\documentstyle[epsfig,subfigure]{mn}
 at 10truept
\title
     [The Cosmological Distribution Function]
{\vglue-3.0truecm
\centerline{\it Accepted for publication in  Monthly Notices}
\vglue 2.5truecm
      Evolution of the cosmological density distribution function      
\author
     [A.N. Taylor \& P.I.R Watts]
     {A.N. Taylor \& P.I.R. Watts \\
     Institute for Astronomy, 
     University of Edinburgh,
     Royal Observatory,
     Blackford Hill, 
     Edinburgh, 
     U.K.\\
	ant@roe.ac.uk,pirw@roe.ac.uk}}
\def\bib{\parskip=0pt\par\noindent\hangindent\parindent
    \parskip =2ex plus .5ex minus .1ex}

\newcommand{\be}{\begin{equation}}
\newcommand{\ee}{\end{equation}}
\newcommand{\ba}{\begin{eqnarray}}
\newcommand{\ea}{\end{eqnarray}}

\newcommand{\fg}{{\mathcal G}}
\newcommand{\nn}{\nonumber \\}
\newcommand{\nnb}{\begin{displaymath}}
\newcommand{\nne}{\end{displaymath}}
\newcommand{\de}{\partial}
\newcommand{\nbar}{\bar{n}}
\newcommand{\x}{\mbox{\boldmath $x$}}
\newcommand{\y}{\mbox{\boldmath $y$}}

\newcommand{\g}{\mbox{\boldmath $g$}}
\newcommand{\E}{\mbox{\boldmath $E$}}
\newcommand{\C}{\mbox{\boldmath $C$}}
\newcommand{\nablab}{\mbox{\boldmath $\nabla$}}
\newcommand{\hMpc}{\,h^{-1}{\rm Mpc}}

\newcommand{\etab}{\mbox{\boldmath $\eta$}}
\newcommand{\rgl}{\rangle}
\newcommand{\lgl}{\langle}

\def\bib{\parskip=0pt\par\noindent\hangindent\parindent
    \parskip =2ex plus .5ex minus .1ex}

\newcommand{\lab}[1]{
	\label{eq: #1 }
}

\begin{document}

\maketitle

\begin{abstract}
We present a new calculation for the evolution of the 
1-point Probability Distribution Function (PDF) of the cosmological 
density field based on an exact statistical treatment. Using the
Chapman--Kolmogorov equation and second--order Eulerian perturbation theory
we propagate the initial density distribution into the nonlinear regime.
Our calculations 
yield the moment generating function, allowing a straightforward
derivation of the skewness of the PDF to second order. We find a 
new dependency on the initial perturbation spectrum.
We compare our results with other approximations to the 1-pt PDF,
and with N-body simulations. We find that our distribution accurately
models the evolution of the 1-pt PDF of dark matter. 
\end{abstract}


\section{introduction}

The standard scenario for the formation of structure in the Universe
is via the gravitational amplification of primordial random Gaussian 
fluctuations 
generated in the Early Universe during an Inflationary phase. An attractive
feature of this scenario is its predictive power in determining the 
history of mass perturbations from initial times up until the present day.
In principle this should allow one to compare observation with theory.
But while detailed predictions at high redshift for the Cosmic Microwave 
Background (CMB) are possible
due to their linearity and dependence on well tested laboratory physics,
predictions at lower redshift are complicated by nonlinear gravitational 
evolution and the physics of galaxy formation. One of the goals of
cosmology is accounting for the statistical evolution of mass and 
galaxies up to the present day.

In this paper we focus on the properties of the one--point density 
distribution function of matter. Interest in the distribution functions
of cosmological fields has grown as they encode a great deal of information
on initial conditions, gravitational evolution and galaxy bias. However
the challenge is to separate out each of these effects. 

The evolution of the 
one--point distribution function has been computed by a number of
authors using a variety of techniques here we present a new approach,
based on propagating the distribution function by the Chapman--Kolmogorov
equation, incorporating nonlinear evolution to second order.
In a further paper we shall develop the formalism to include the effects 
of galaxy biasing and redshift--space distortions (Watts \& Taylor 2000).


The 1-pt PDF is a 
useful quantity in cosmology. While reasonably straightforward to 
measure (see Hamilton 1985, Alimi et al 1990, Szapudi, Szalay \& Bosch\'{a}n 
1992, Gazta\~{n}aga 1992, 1994, Bouchet et al 1993, Szapudi, Meiksin \& Nichol 
1996, Kim \& Strauss 1998) the 1-pt PDF embodies the full hierarchy of 
correlation functions and its measurement ensures that physical constraints 
such as the Lyapunov inequalities for the moments 
($\lgl x^{a+b} \rgl \ge \lgl x^a \rgl \lgl x^b \rgl$,
for any random variable $x$) are automatically satisfied. This is not
necessarily the case for direct measurement of the moments. In addition
the PDF offers a convenient way of dealing with Poisson sampling in the 
galaxy distribution.

From a theoretical perspective, the 1-pt PDF is a convenient quantity
to calculate if it is initially Gaussian. A number of different methods
have been developed to calculate its evolution.
Fry (1985) first suggested calculating the probability function by
applying a hierarchical series suggested by the BBGKY equations to
solve for the moment generating function.
Bernardeau (1992) derived an exact expression for the 
evolution of the moment generating function. The generalisation to 
top-hat filtered 
density and velocity fields is found in Bernardeau (1994), while
Bernardeau (1996) studied the 2-point cumulants and the PDF. However
so far a generalisation of these results to include bias and redshift--space
distortions in the PDF has not been presented.

An approximation for the PDF can also be constructed from the first few
cumulants, in the limit of small variance, by the Edgeworth expansion
(Juszkiewicz et al., 1995 and Bernardeau \& Kofman, 1995), where the
cumulants have been derived directly from Eulerian perturbation theory
(Bouchet et al., 1992) or Lagrangian perturbation theory (Bouchet et al.
1995). This can be extended to redshift--space using the Lagrangian 
perturbation calculations of Hivon et al. (1995) for the skewness.
Colombi et al. (1997) introduced an extended perturbation theory, where
the results of perturbation theory are allowed extra freedom and extrapolated
to the nonlinear regime.

As well as Eulerian perturbation theory other approximations have also
been applied. Kofman et al (1994) and Bernardeau \& Kofman (1995) derived
the PDF in  the Zel'dovich approximation, while Hui, Kofman \& Shandarin (1999)
extended this to include the effects of redshift--space distortions.

A more phenomenological approach was taken by 
Coles \& Jones (1991) who approximated the properties of the 
PDF by a lognormal distribution while Colombi (1994) suggested an Edgeworth 
expansion about the lognormal distribution.



In this letter we apply second--order perturbation theory to an initially 
Gaussian distributed density field to calculate the exact second--order 
characteristic function. We then numerically inverse Fourier transform this
to yield the 1-pt pdf. Our method therefore treats the propagation of 
probabilities exactly.

The letter is organised as follows. In Section 2 we calculate the
evolution of the cosmological probability distribution function and
discuss the transformation to a discrete count distribution.  In
Section 3 we compare our results with the results of N-body
simulations and with other distributions in Section 4. Conclusions are
presented in Section 5. We begin by describing second--order
perturbation theory and calculating the probability distribution
function.

\section{The cosmological probability  distribution function}
\label{sec-rspdf}

\subsection{Second--order perturbation theory}

Before shell--crossing, and in a spatially flat universe, the Eulerian 
density field, $\delta(\x,t)$, can 
be expanded in a series of separable functions;
\be
	\delta(\x,t)=\sum_{n=1}^\infty \delta_n(\x,t)
		    =\sum_{n=1}^\infty D_n(t) \varepsilon_n(\x),
\ee
where $\varepsilon_n$ is an $n^{th}$-order time-independent density 
field and $D_n$ is an
$n^{th}$-order universal growth function. It is a special property of this
expansion that the spatial and time components are separable. In 
a universe with spatial curvature this expansion is only possible to 
second--order.

The growth of perturbations in the nonlinear regime can be 
calculated by solving the continuity, Euler and Poisson equations 
sequentially for each order in the perturbation expansion.
To second--order the density field can be derived from linear quantities
by the relation
(Peebles 1980, Bouchet et al 1992)
\be
\delta=\delta_{1}+\frac{1}{3}(2-\kappa)\delta_1^2
 -\etab . \g+\frac{1}{2}(1+\kappa)E^{2},
\label{nl}
\ee 
where 
\be
	\etab(\x,t) = \nablab \delta_1(\x,t)
\ee
is the gradient of the linear density field and
\be
	\g(\x,t) = - \nablab \nabla^{-2} \delta_1(\x,t)
\ee
is the linear peculiar gravity field\footnote{In this paper we define
 $4 \pi G \rho_0=3/2\Omega H^{2} = 1$ and the expansion parameter $a(t)=1$, 
since our final distribution function will be dimensionless.} 
where $\nabla^{-2}$ is the inverse Laplacian. The trace-free tidal 
tensor is given by 
\be
	E_{ij}(\x,t) = \nabla_i \nabla_j \nabla^{-2} \delta_1(\x,t) - 
	\frac{1}{3} \delta_1(\x,t) \delta_{ij} .
\ee
Equation (\ref{nl}) is the general result for an arbitrary cosmology
where\footnote{We define $\kappa$ following Bouchet et al. (1995) and 
Hivon et al. (1995), but note that this differs from the definition of 
Bouchet et al. (1992) who define $\kappa\approx2/14 \Omega^{-2/63}$}  
$\kappa=D_2/D^2_1\approx -3/7 \Omega^{-2/63}$ (Bouchet et al 1992,
see also Catelan et al. 1995), and
$\delta_1(\x,t)=D_1(t)\varepsilon_1(\x)$, where 
$D_1(t)\approx(1+z)^{-\Omega^{0.6}}$ is the linear growth function 
(Peebles 1980).

\subsection{The distribution function of initial fields}

In order to calculate the second--order density distribution function,
$P(\delta)$, we must first
calculate the joint probability of each of the fields in equation
(\ref{nl}), $P(\delta_{1},\etab,\g,\E)$. Defining the parameter
vector $\y=(\delta_1,\etab,\g,\E)$ the joint distribution function is 
given by the multivariate Gaussian
\be
P(\y)=\frac{1}{((2\pi)^{n}\mid\det{\C}\mid)^{1/2}} \exp{\left
(-\frac{1}{2} \y^t\C^{-1}\y
\right)},
\label{mult}
\ee
where 
\be
	\C = \lgl \y \y^t \rgl
\ee
 is the $12\times12$ covariance matrix that gives the degree of
correlation between each of the components of $\y$. 
The elements of the covariance matrix are 
\ba
	\lgl \delta^2 \rgl &=& \sigma^2_0, \hspace{1.3cm}
	\lgl \eta_i \eta_j \rgl = \frac{1}{3}\sigma^2_1 \delta_{ij}, \nn
	\lgl g_i g_j \rgl &=&  \frac{1}{3}\sigma^{2}_{-1} \delta_{ij}, 
	\hspace{0.5cm}
	\lgl \eta_i g_j \rgl = \frac{1}{3}\sigma^2_0 \delta_{ij}, \nn
	\lgl E_{ij} E_{kl} \rgl &=& \frac{1}{15} \sigma^2_0 
	\left(\delta_{ik}\delta_{jl}+\delta_{il}\delta_{jk}-
	\frac{2}{3}\delta_{ij}\delta_{kl}\right),
\ea
with the remaining entries in the covariance matrix zero.
These variances are defined as
\be 
	\sigma_{n}^{2} = D^2(t) \int_0^\infty \! \frac{dk}{2 \pi^2} 
			\, k^{2+2n}  P(k) ,
\ee
where $P(k)$ is the initial power spectrum of density perturbations.
The components of this distribution are
\be
\y^t\C^{-1}\y = \frac{\delta^{2}}{\sigma_{0}^{2}} +
		\frac{3}{(1-\gamma_{\nu}^{2})}
		\left(\frac{\eta^{2}}{\sigma_{1}^{2}} +
		\frac{g^{2}}{\sigma_{-1}^{2}} -
		2\gamma_{\nu} \frac{\etab . \g}{\sigma_1 \sigma_{-1}} \right) 
  		+  15 \frac{E^2}{\sigma_{0}^{2}}
\ee
and
\be
\det \C = \frac{20}{3105^3\pi^4}\sigma_0^{12}\sigma_1^6
	\sigma_{-1}^6(1-\gamma_\nu^2)^3.
\ee
 The correlation parameter, $\gamma_{\nu}$, is defined as 
\be
\gamma_{\nu} = \frac{\sigma_{0}^{2}}{\sigma_{1}\ \sigma_{-1}},
\ee
providing a measure of the correlation between the linear
velocity and density gradient fields. 
If we assume a power-law power spectra with a Gaussian cut-off, 
$P(k)\propto k^{n}e^{-k^2 R^2}$, where n is the 
spectral index and $R$ some arbitrary length scale, then 
\be
\gamma_{\nu}=\frac{n+1}{n+3},
\ee
with the constraint $n > -1$ to insure convergence of the velocity field.
The correlation parameter must be positive definite since under gravity matter 
will be displaced from  low-- to high--density regions. The special 
case of $n \rightarrow -1$ for a power-law
initial spectra results in $\gamma_\nu=0$ because the velocity field
diverges on small scales. In fact the diverging velocities form an
incoherent Gaussian random field which is uncorrelated with the 
density field (Taylor \& Hamilton 1996).

\subsection{Propagation of the density distribution function}

The distribution function can be propagated to later times by 
the Chapman--Kolmogorov equation,
\be
	P(\x) = \int  d \y \, W(\x|\y) P(\y)
\label{chapkol}
\ee
where $W(\x|\y)$ is the transition probability from $\x$ to $\y$. In the
case of a deterministic process, such as the one considered here, the
transition probability reduces to a delta function restricting the 
number of possible paths of evolution to one. This transition probability
is given by
\be
	W(\delta|\y) = \delta_D\big[\delta-\delta_{1}-
\frac{1}{3}(2-\kappa)\delta_1^2
 +\etab . \g - \frac{1}{2}(1+\kappa)E^{2}\big]
\ee
Inserting this into equation (\ref{chapkol}) we find 
\be
P(\delta)=\langle\delta_{D}(\delta-\delta(\y))\rangle_{y}.
\label{expt}
\ee
where $\delta(\y)$ is the right hand side of  equation
(\ref{nl}). Hence for deterministic transitions the Chapman--Kolmogorov 
equation becomes the expectation value of the delta function. This is a 
well--known result from probability theory (eg van Kampen 1992).

Fourier transforming the delta function we find
\be
	P(\delta)\equiv\frac{1}{2\pi}\int_{-\infty}^\infty dJ \, 
	\fg(J) \exp (iJ\delta)
\label{gen1}
\ee 
where 
\ba
	\fg(J) &\equiv&\int_{-\infty}^\infty
	 	d\delta P(\delta)\exp \left(-iJ\delta\right) \nn
	       &=& \lgl \exp \left(-iJ\delta(\y)\right) \rgl_y
\label{gen2}
\ea
is the characteristic function. In the second line
we have used equation (\ref{expt}) to write the 
characteristic function as an expectation over all
the linear fields, $\y$. This expression reduces to a set of
multivariate Gaussian--type integrals that we can easily evaluate
yielding
\be
	\fg(J)=\Theta(J)
	\exp{\left[-\frac{J^{2}\sigma_{0}^{2}}{2(1+i\alpha_{1}J)}\right]}
\label{chrfn1}
\ee
where
\be
	\Theta(J)   =  (1+i\alpha_{1}J)^{-1/2}(1+ i\alpha_{2}J)^{-5/2}
		(1-i\alpha_{3}J + \alpha_{4}J^{2})^{-3/2}
\label{chrfn2}
\ee
and where the $\alpha$ coefficients are given by
\begin{displaymath}
\alpha_{1}=\frac{2}{3}(2-\kappa)\sigma_{0}^{2}, \ \ 
\alpha_{2}=\frac{2}{15}(1+\kappa)\sigma_{0}^{2}, \ \ 
\end{displaymath}
\be
\alpha_{3}=\frac{2}{3}\sigma_{0}^{2},   \ \ 
\alpha_{4}=\frac{1}{9}\frac{(1-\gamma_{\nu}^{2})}{\gamma_{\nu}^{2}}
\sigma_{0}^{4}
\label{def}
\ee
The probability distribution can then be found by numerically
integrating equation (\ref{gen1}). Equations (\ref{chrfn1}), (\ref{chrfn2})
and (\ref{def}) are the central results of this letter.

The characteristic function for $\delta$ can be separated into characteristic
functions for each term in the summation in equation (\ref{nl}). This is
to be expected, since the distribution of a summation of random variables 
is equal to the convolution of their individual distributions, and so 
the characteristic functions just multiply. We expect this to be true
for all orders of the perturbation series. But rather than converging to
a Gaussian distribution, as suggested by a naive application of the
Central Limit Theorem, the final distribution is driven away from it by 
the correlations induced by gravity.

\begin{figure}
\centering
\subfigure{\epsfig{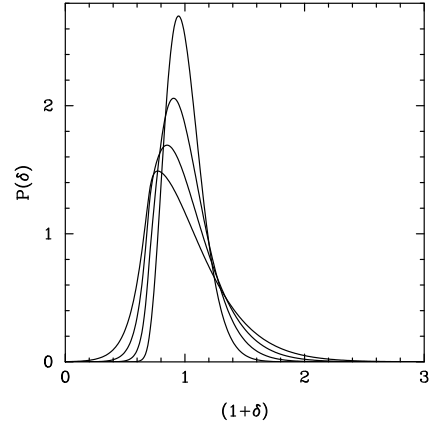}}
\subfigure{\epsfig{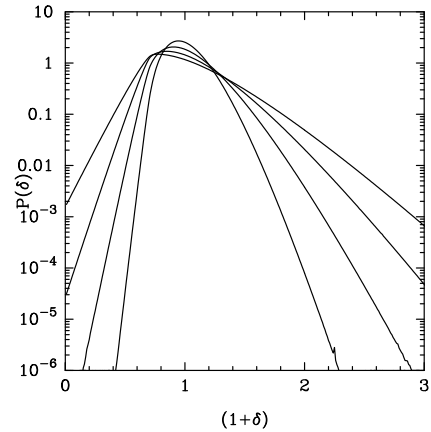}}
\caption{The evolution of the 1-pt distribution function, $P(\delta)$ 
calculated from second--order perturbation theory. The variances are
$\sigma_{0}=0.15,\ 0.20,\ 0.25,\ 0.3$. In the lower plot we use 
logarithmic axis to emphasis the tails of the distribution.}
\end{figure}

Figure 1 shows the evolved density distribution function 
for a range of $\sigma_{0}$.
We use linear axis for the top plot emphasising the peak and 
logarithmic axis in
the lower plot emphasising the tail of the distribution. As expected, 
the shape of the distribution is very nearly
Gaussian when the variance is small, becoming very
rapidly non-Gaussian for higher values. For high variances the 
probability density does not drop to zero at $\delta=-1$, since in 
second--order perturbation theory the density field can be negative,
generating non-vanishing regions with $\delta \leq -1$. This is also 
true of linear theory where there are always negative density regions
for Gaussian initial conditions. This is a feature of Eulerian 
distribution functions. Those calculated in Lagrangian perturbation
theory, or the lognormal, have positive definite densities at all times.

\begin{figure}
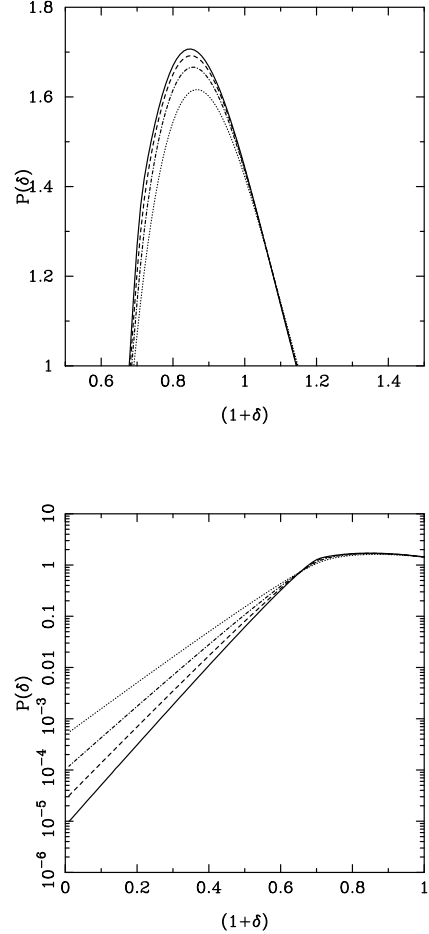

\centering
\subfigure{\epsfig{figure=pdf2a.eps,width=6.0cm,angle=270,clip=}}
\subfigure{\epsfig{figure=pdf2b.eps,width=6.0cm,angle=270,clip=}}
\caption{The variation in the peak (upper) and tail (lower) of the
1-point density distribution function as a function of the correlation
parameter, $\gamma_\nu$.  The values used were $\gamma_\nu=0.65$
(solid line), $0.55$ (dashed), $0.45$ (dot-dashed) and $0.35$
(dotted). The effect is small for CDM-type initial power spectra where
the correlation parameter is in the range $0.55<\gamma_\nu<0.65$ over
a wide range of scales.}
\label{figgam}
\end{figure}

In Figure \ref{figgam} we demonstrate the effects of the correlation parameter,
$\gamma_\nu$, on the PDF. The main effects are an increase in volume of
underdense regions with a corresponding decrease in extremely underdense
regions when $\gamma_\nu$ is high. For low $\gamma_\nu$, the underdensities
are smaller and deeper. The physical reason for this is that the $\etab .\g$
term in second--order perturbation theory deals with the evacuation
of the voids, rather than the amplification of peaks. When $\gamma_\nu$
is low this term is weakened and voids tend to be smaller and deeper,
as they would be if the linear field were extrapolated.  Increasing this
correlation widens the voids, but makes them shallower to help satisfy the 
$\delta \ge -1$ constraint. This effect is small for CDM-type initial 
power spectra where the
correlation parameter is in the range $0.55<\gamma_\nu<0.65$ over a wide 
range of scales.

\subsection{Skewness from the characteristic function}

Since our characteristic function is correct to second--order a
useful check is to calculate the variance and skewness to compare
with previous estimates.
Taking the derivatives of the characteristic function with respect to $J$
and setting $J$ equal to zero yields the moments of the evolved 
density distribution function,
\be
	 \frac{\partial^{n}}{\partial [iJ]^{n}} \fg(J=0)=
	\langle \delta^{n} \rangle.
\lab{genfunc}
\ee
In the literature it is common to express the moments in the form of
the moment parameters 
\be
	S_n=\langle \delta^{n} \rangle_{c}/\langle
	\delta^{2}\rangle^{n-2}
\ee
 where $\langle \delta^{n}\rangle_{c}$ is the
connected or irreducible part of $\langle \delta^{n} \rangle$. 
The irreducible moments, or cumulants, can be generated
by 
\be
 \frac{\partial^{n}}{\partial [iJ]^{n}} \ln \fg(J=0)=\langle 
\delta^{n} \rangle_c
\lab{irrgenfunc}
\ee
From this we can calculate the second and third order connected moments 
of our distribution function; 
\be
	S_2=1
\ee
and
\be
	S_{3} = 2(2-\kappa) \stackrel{\Omega\rightarrow 1}{
	\longrightarrow} \frac{34}{7}
\ee
to lowest order, reproducing the results of Peebles (1980).
Hence our PDF leads to the correct second--order skewness. 
The intrinsic effects of the shape of the
power spectrum via the correlation coefficient $\gamma_v$ are to 
higher order. 


\subsection{The discrete distribution function}

In reality the density field of galaxies is not a continuous function,
but a discrete distribution. To account for this it is usual to assume
Poisson sampling of the continuous density field as a crude approximation
to galaxy formation processes. Any variation is attributed to biased
galaxy formation, which modulates the underlying function. We shall 
consider this elsewhere (Watts \& Taylor, 2000). 

The continuous density distribution function can be transformed 
to a discrete form by the expectation
\be
	P(n)=\frac{\nbar^n}{n!} 
	\lgl (1+\delta)^n e^{\nbar(1+\delta)}\rgl_\delta,
\ee
where $\nbar$ is the mean galaxy count and 
the expectation is taken with respect to the nonlinear density distribution.
Expanding this in terms of the characteristic function and taking the
expectation we find
\be
P(n)=\int_{-\infty}^{\infty}dJ\
G(J)\left(1-\frac{iJ}{\nbar}\right)^{-n-1}\exp{(-iJ)}
\ee
In the limit $\nbar \rightarrow \infty$ this 
returns the continuum distribution where $\delta=n/\nbar-1$.
We note that again that the discrete distribution has a non-vanishing value
at $n=0$, $P(n=0)$, since  the probability of finding a no galaxies
within a cell is finite. This is the Void Probability Function (White 1979).

It is also useful to have the generating function for the discrete
moments of this distribution, $G_n(k)$. Following some straightforward 
calculation we find that 
\be
	G_n(k) = G[ i \nbar(e^{-ik}-1)] \exp\nbar(e^{-ik}-1).
\ee
The discrete moments can then be found by differentiating $G_n(k)$;
\be
	\lgl \delta^m \rgl = \frac{\de^m G_n(k=0)}{\de (ik)^m}
\ee
Again the connected moments can be found by differentiating 
$\ln G(k)$ with respect to $ik$.

\section{Comparison of results with n-body simulations}

In this section we show a comparison between our theoretical PDF and
the counts in cells PDF found from cosmological n-body simulations. We
used a version of Hugh Couchman's (1991) Adaptive P$^3$M code altered by 
Peacock \& Dodds (1994) to allow simulations of low density open and
flat universes.

The simulation volume was a periodic cube of comoving 
side $200 \, h^{-1}{\rm Mpc}$ 
containing $100^{3}$ particles. We chose a CDM--type linear power
spectrum of Gaussian initial perturbations (Bardeen et al. 1986), 
normalised to match the
observed abundance of clusters with linear variance given by 
$\sigma_8=0.6\Omega^{-0.53}$ (Viana \& Liddle 1996). The 
simulation was carried out on a $128^{3}$ Fourier mesh.

To measure the PDF from the numerical simulations  we
smoothed the discrete particle distribution with a Gaussian filter of
radius $R$. The PDF was then found from the smoothed density field
evaluated on a $128^{3}$ grid. Since the effective radius of the
binning grid was much smaller than that of the filter radius we
expected negligible contribution to the smoothing from it.

\begin{figure}
\centering
\subfigure{\epsfig{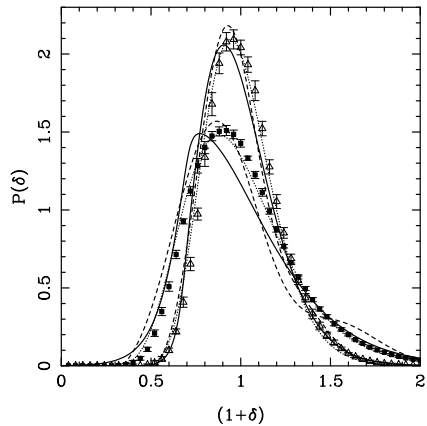}}
\subfigure{\epsfig{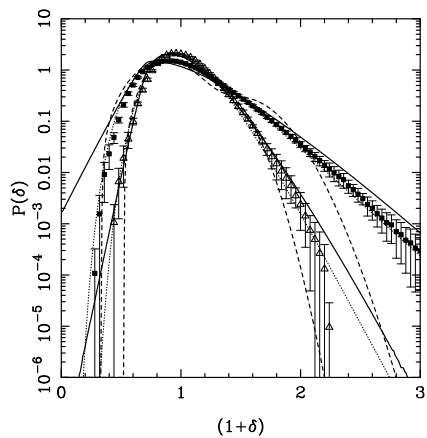}}
\caption{Comparison of the second--order PDF (solid line) with the 
results of N-body simulations (points). The variances are $\sigma_0=0.2$ (open
triangles) and $0.3$ (filled squares). Also plotted are the corresponding
distribution functions for the lognormal model (dotted) and Edgeworth
expansion (dashed).}
\label{numsim}
\end{figure}

The choice of variance to use in our model is slightly ambiguous,
given that we do not treat smoothing exactly. Hence it is better 
to match variances rather than smoothing scales. However the variance in 
second--order perturbation theory is the same as in 
linear theory so we could use either the linear input variance from 
the simulations, or take the measured nonlinear variance when 
making a comparison. In practice we found that all of the models provided a
better fit if the nonlinear variances where used.
This is a well know effect and 
is the basis of the extended perturbation theory approach of 
Colombi et al (1997).

Figures (\ref{numsim}, top) and (\ref{numsim}, bottom) show the results of 
numerical simulations (points) alongside our second--order
PDF (solid line), the lognormal model (Coles \& Jones 1991, dotted)
and the Edgeworth expansion (Juszkiewicz et al. 1995, 
Bernardeau \& Kofman 1995; dashed).  
Each plot shows the PDF for two different variances, $\sigma_0=0.2$,
smoothed on a scale of $R=20\hMpc$
(open circles) and $0.3$ (filled squares) smoothed on $13.5\hMpc$.

The effects of shot noise were taken into account as shown in Section
2.3 with the mean particle density $\bar{n}$ chosen to coincide with
the mean particle count in a boxes with side $l=\sqrt{5}R$. Since $\bar{n}$
was a very large number in all cases, shot noise effects were
extremely small. This would not be the case in a real galaxy survey
with lower number counts. The remaining factor controlling the shape
of the theoretical PDF was $\gamma_{\nu}$. For linear CDM power
spectra the calculated $\gamma_{\nu}$ was approximately 0.55 for a
wide range of smoothing radii and it was this value that was used to
prepare all the plots in this section.

Error bars on the N-body data points are the standard deviation over 5
independent simulations with identical cosmological parameters. 

The linear axes of figures (\ref{numsim}, top) shows the accuracy around
the peak of the distributions while the logarithmic scale of figure 
(\ref{numsim}, bottom) shows the tails of the distributions. At low
variance ($\sigma_0=0.2$) we find very good agreement between all of
the models and the simulations. Overall we found that the lognormal 
model fitted the simulations extremely well for all values of the 
variance and around both the peak and the tails. As has been 
remarked elsewhere we regard this as something of a fluke (Bernardeau \&
Kofman 1995), and that the 
lognormal makes a very useful fitting function.

Our second--order PDF fits the peak fairly accurately, but tends to be
slightly too skewed. This effect becomes greater at higher variance.
The tails of the distribution match the N-body simulations rather
well, although at high variances and low densities the distribution is
too high. This is due to the real distribution being constrained to go
to zero at zero density, whereas, as we have already remarked the
second--order PDF is not. In comparison the Edgeworth expansion, taken
to second order (Juszkiewicz et al. 1995, Bernardeau \& Kofman 1995)
\be 
	P_E(\delta)=\left[1+\frac{1}{6}S_3 \sigma_0
	H_3(\delta/\sigma_0)\right] 
	\frac{e^{-\delta^2/2\sigma_0^2}}{\sqrt{2\pi}}, 
\ee 
where $H_n(x)$ is a Hermite polynomial, is also slightly too skewed
and tends to undershoot the high density tail compared to the
simulations. At large variance the expansion breaks down and the
Edgeworth distribution develops a wiggle on the high density tail. In
fact the Edgeworth expansion fares badly even when taken to third
order because the unphysical wiggles are exaggerated and the low
density tail becomes more rapidly negative.

\section{Comparison with other approximations}

\begin{figure}
\centering
\subfigure{\epsfig{figure=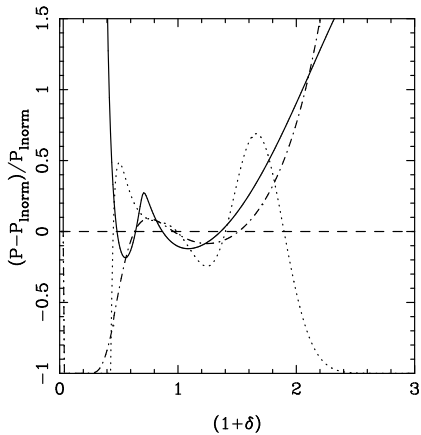,width=6.0cm,angle=270,clip=}}
\caption{Comparison with other distributions. We normalise all the
models to the lognormal distribution ($P_{\rm lnorm}(\delta)$; dotted
line), since we find that this best fits our N-body simulation results
over a wide range of parameter values. The plot shows the ratio
$(P(\delta)-P_{\rm lnorm}(\delta))/P_{\rm lnorm}(\delta)$ for our
second--order calculation (solid line), the Zel'dovich approximation
(Kofman et al. 1994; dot-dashed line) and the Edgeworth expansion
(dotted line). The variance used for the comparison was
$\sigma_0=0.25$.}
\label{figcomp}
\end{figure}

Finally we compare our PDF with other approximations for the 
1-point density distribution. In figure \ref{figcomp} we show 
three distributions, our second--order PDF calculation (solid line), the
Zel'dovich approximation (Kofman et al. 1994; dot-dashed line) and the
Edgeworth expansion (dotted line) plotted relative to the lognormal 
distribution. Our choice of the lognormal distribution as the point
of normalisation is based on the agreement we find between it and
our N-body simulations.

Over the range $0.4<\delta<1.5$ all of the models agree to within a
few percent, with the exception of the second--order Edgeworth expansion. 
But for densities below this regime, our second--order PDF
overpredicts because the positive density condition is only weakly 
met. Both the Zel'dovich and Edgeworth models underpredict the
probability of low density regions. Above the regime of good
agreement our model again overpredicts, again because we over-extrapolate
the high density evolution, but not as much as the Zel'dovich approximation,
where caustic formation tends towards a too high distribution.
In contrast to both models the Edgeworth approximation underpredicts 
the high density regions and develops a large wiggle in the positive 
density regime.

\section{Summary}

We have presented a new method for calculating the nonlinear evolution
of the 1-point density distribution function using the
Chapman--Kolmogorov equation to propagate the probability
distribution, and second--order perturbation theory to evolve the
density field. This has the advantage over other methods that the
resultant probability distribution must be positive definite, and can
be readily extended to include the effects of Eulerian deterministic
or stochastic bias and redshift space distortions.  The main
disadvantages of our method are that it is not obvious how to include
the effects of smoothing in the final distribution and it is
difficult to see how one could extend the evolution of the density
field to higher order. However, despite these problems we find
extremely good agreement with numerical simulations and that our
distribution is an improvement on other approximations. Since we
derive the characteristic function the moments and cumulants
(connected moments) of the field are easily derived, as is the
distribution of any local transformations of the density field. The
formalism we have introduced can also be used to calculate more
complicated \mbox{1- and 2-point} distributions of cosmologically
interesting fields.

\section*{Acknowledgements}
PIRW thanks the PPARC for a postgraduate grant. ANT thanks the PPARC
for a postdoctoral fellowship.


\section*{References}

\bib Alimi J.-M., Blanchard A., Schaeffer R., 1990, ApJ, 349, L5

\bib Bardeen J.M., Bond J.R., Kaiser N., Szalay A.S., 1986, ApJ, 304, 15

\bib Bouchet F.R., Juszkiewicz R., Colombi S., Pellat R., 1992, ApJ, 394, L5

\bib Bouchet F.R., Strauss M.A., Davis M., Fisher K.B., Yahil A., Huchra J.P.,
	1993, ApJ, 417, 36

\bib Bouchet F.R., Colombi S., Hivon E., Juszkiewicz R., 1995, AA, 296, 575 

\bib Bernardeau F., 1992, ApJ, 392, 1

\bib Bernardeau F., 1994, AA 291, 697

\bib Bernardeau F., 1996, AA, 312, 11

\bib Bernardeau F., Kofman L., 1995, ApJ, 443, 479

\bib Catelan P, Lucchin F., Matarrese S., Moscardini L, 1995, MNRAS, 276, 39

\bib Coles P., Jones B., 1991, MNRAS, 248, 1

\bib Colombi S., 1994, ApJ, 435, 536

\bib Colombi S., Bernardeau F., Bouchet F.R., Hernquist L., 1997, 
	MNRAS, 287, 241

\bib Couchmann 1991, ApJ, 368, L23


\bib Fry J.N., 1985, ApJ, 289, 10

\bib Gazta\~{n}aga E., 1992, ApJ, 398, L17

\bib Gazta\~{n}aga E., 1994, MNRAS, 286, 913

\bib Hamilton A.J.S., 1985, ApJ, 292, L35

\bib Hivon E., Bouchet F.R., Colombi S., Juszkiewicz R., 1995, AA, 298, 643

\bib Hui L., Kofman L., Shandarin S.F., 1999 (astro-ph/9901104)

\bib Juszkeiwicz R., Weinberg D.H., Amsterdamski P., 
	Chodorowski M., Bouchet F., 1995, ApJ, 442, 39

\bib Kim R.S.J., Strauss M.A., 1998, ApJ, 493, 39

\bib Kofman L., Bertschinger E., Gelb J.M., Nusser A., Dekel A., 
1994, ApJ, 420, 44

\bib Viana P.T.P., Liddle A.R., 1996, MNRAS, 281, 323

\bib Peacock J.A., Dodds S., 1994, MNRAS, 267, 1020

\bib Peebles P.J., 1980, ``Large--Scale Structure in the Universe'', 
	Princeton University Press, Princeton

\bib Szapudi I., Szalay A., Bosch\'{a}n P., 1992, ApJ, 390, 350

\bib Szapudi I., Meiksin A.,  Nichol R., 1996, ApJ, 473, 15

\bib Taylor A.N., Hamilton A.J.S., 1996, MNRAS, 282, 767

\bib van Kampen, N.G., 
1992, ``Stochastic Processes in Physics and Chemistry'',
	North--Holland, Amsterdam

\bib Watts P.I.R, Taylor A.N., 2000, in Preparation

\bib White S.D.M, 1979, MNRAS, 186, 145

\end{document}